# A convolutional neural network of low complexity for tumor anomaly detection

Vasileios E. Papageorgiou[1][0000-0002-8131-3484], Pantelis Dogoulis[2][0000-0003-3848-9927] and Dimitrios-Panagiotis Papageorgiou[3][0000-0002-3867-7135]

[1] Aristotle University of Thessaloniki, Thessaloniki, Greece
vpapageor@math.auth.gr
[2] Center of Research and Technology Hellas, Thessaloniki, Greece
dogoulis@iti.gr
[3] Aristotle University of Thessaloniki, Thessaloniki, Greece
depapage@math.auth.gr

**Abstract.** The automated detection of cancerous tumors has attracted interest mainly during the last decade, due to the necessity of early and efficient diagnosis that will lead to the most effective possible treatment of the impending risk. Several machine learning and artificial intelligence methodologies has been employed aiming to provide trustworthy helping tools that will contribute efficiently to this attempt. In this article, we present a low-complexity convolutional neural network architecture for tumor classification enhanced by a robust image augmentation methodology. The effectiveness of the presented deep learning model has been investigated based on 3 datasets containing brain, kidney and lung images, showing remarkable diagnostic efficiency with classification accuracies of 99.33%, 100% and 99.7% for the 3 datasets respectively. The impact of the augmentation preprocessing step has also been extensively examined using 4 evaluation measures. The proposed low-complexity scheme, in contrast to other models in the literature, renders our model quite robust to cases of overfitting that typically accompany small datasets frequently encountered in medical classification challenges. Finally, the model can be easily re-trained in case additional volume images are included, as its simplistic architecture does not impose a significant computational burden.

**Keywords:** Convolutional neural networks; Tumor detection; Biomedical image classification; Data augmentation; Entropy; Artificial intelligence.

## 1 Introduction

Applications of artificial intelligence (AI) in medicine continue to grow and affect every aspect of cancer care. These applications fall into 2 main categories, namely supervised and unsupervised learning [1-2]. In supervised learning, computers learn and adapt by studying labeled biomedical instances to copy the diagnostic skills of experienced oncologists. Three of the most common and deadly tumors that affect people's quality of life in everyday life are brain, lung, and kidney tumors, making their early detection an important concern.

A brain tumor is an abnormal mass of tissue where cells grow uncontrollably. Based on the World Health Organization (WHO), these types of tumors account for less than

2% of human cancers, although their severe morbidity and associated complications make timely diagnosis an important concept in the field of medicine [3]. Intracranial tumors can be fatal, worsen the patient's standard of living, and can affect men, women, or children.

Lung cancer ranks second, accounting for approximately 11.4% of total cancer cases, with an estimated 2.2 million lung cancer cases only. Lung tumors represent the leading cause of death due to cancer, with deaths accounting for 18% of all cancers [4]. The survival rate of patients suffering from lung tumors 5 years after diagnosis ranges between 10-20%. Screening with low-dose computed tomography (CT) could aid the timely detection of lung tumors so that the disease can be possibly treated. In addition, it has been reported that the patient's likelihood of living a long life increase if the tumor is timely diagnosed and treated successfully [5].

Renal cell carcinoma (RCC) is the 6th most ordinary cancer among all tumors in men and the 10th most common in women. Despite advances in understanding the molecular biology of RCC and refinements in therapies, treating patients with RCC at any stage of the disease is challenging. Detection of early-stage renal tumor has improved in recent decades with the use of cross-sectional imaging [6]. Most renal carcinomas are initially detected as incidental renal masses on cross-sectional imaging performed for unspecified disease. Although most are discovered as small renal masses (70%), earlier definitive therapeutic intervention for these tumorous regions has not led in a notable advancement in cancer mortality [7].

In our analysis, we present convolutional neural network (CNN) architecture of low complexity for tumor anomaly diagnosis. The reduced complexity of the proposed network makes it suitable for studying small datasets which often accompany medical analyses, compared to many articles that employ complex and computationally expensive schemes. This architecture not only eliminates the likelihood of overfitting – which can be a major issue in both statistical and AI models [8] – but also increases the flexibility and adaptability of the approach, as it can be re-trained without significant computational cost in the occasion where new MRI or CT images are added to the dataset. In addition, various data augmentation techniques have been implemented and their respective impact on classification efficiency has been studied in detail.

In contrast to most studies included in the literature, our novel architecture has been tested on not only one, but 3 tumor datasets. These datasets contain normal, benign and malignant instances corresponding to brain, lung and kidney tumors. Hence, the variety of different datasets and the diversity of cancer classification problems, validate the trustworthiness of the produced detection efficiency, rendering our approach a reliable medical tool that can be easily utilized in modern oncology.

## 2 Related Work

Regarding medical imaging, there are a variety of algorithms from the field of AI for brain cancer classification/detection. Characteristic examples are support vector machines (SVM) [9], K-nearest neighbors (KNN) [10] and artificial neural networks (ANN) [11]. In parallel, CNNs seem to be the most suitable for processing MRI or CT images due to their high classification performance. There are several articles that address binary and multiclass tumor challenges, utilizing a number of state-of-the-art deep CNNs. The most relevant articles regarding our analysis corresponds to binary classification attempts of MRI and CT images displaying brain, lung or kidney tumors.

The authors in [12-14] combine image augmentation and preprocessing methods with conventional CNN methods aiming to classify benign and malignant tumors with accuracy of 97.5%, 94.1% and 98% respectively. In [15-17], hybrid CNN-SVM models are used for binary classification with corresponding accuracies of 88.54%, 95%, and 95.62%. The authors in [16] and [17] propose the usage of ResNet for the part of the feature extraction and a SVM for the feature extraction. The examined tumor images were preprocessed by entropy segmentation techniques and resolution enhancement.

Regarding lung tumors, in [18-19] the authors use SVM classifiers and CNN GoogleNet based on the IQ-OTHNCCD dataset. In [18], the authors preprocess the images from CT using Gaussian filters, bit-plane slicing and image segmentation and achieve an accuracy of 89.88%, while in [19] they implement the Gabor filter and regions of interest (ROI) extraction with a grouping accuracy of 94.38%. Other machine learning attempts for lung cancer classification include the utilization of KNN [20], SVM [21], Naive Bayes [20], and Random Forests [22]. In [23], the authors implement CNN architectures such as VGG16, MobileNet, AlexNet, DenseNet, VGG19, and ResNet with efficiency between 48-56% for classification of normal, benign, and malignant tumors. Polat and Mehr [24] utilize a hybrid 3D CNN-SVM with a classification efficiency of 91.81%.

Finally, several articles have addressed renal cancer detection, particularly using preprocessed forms of CT images. In machine learning approaches to discriminate between benign and malignant tumors [25-27], most studies focused on discriminating between low- fat angiomyolipomas and renal cancer with promising results, with the AUC metric reaching values between 0.90 and 0.96. Han et al [28], using a modified GoogleNet architecture, encountered lower performance on a 3-class problem for identifying papillary RCCs (pRCCs) compared to clear cell RCCs (ccRCCs) and chromophobe RCCs (chrRCCs), compared to the binary classification problem of ccRCCs compared to non-ccRCCs. Furthermore, Li et al [29] identified low- and high-grade ccRCC, respectively, based on MRI combined with patient history and radiologist-assigned imaging features and achieved an AUC of 0.845.

## 3    Methodology

In this part, we describe the complete framework that was constructed in order to train our proposed architecture and boost its generalization ability. In more detail, we analyze the source of each dataset and the classes that are contained. Then we provide the proposed low-complexity convolutional neural network and finally we describe the data augmentation method that was applied. We also describe in detail the main concepts of the CNNs as well as the metrics that are used in the inference stage.

## 3.1 Convolutional Neural Networks

Convolutional Neural Networks is a class of ANN architectures that are most commonly based on the convolution kernels. These networks are mainly used in visual related problems such as video classification, image segmentation and medical image analysis. A CNN includes the basic modules of *convolutional layers* (which are the most important parts of feature extraction), *pooling layers* (which are utilized in order to reduce the dimensionality of the preprocessed matrices), *batch normalization layers* (which are helpful in the computational stability during training) as well as the *fully connected layers* (which is used as the feature selection mechanism).

Convolutional layers consist of kernel sets representing the model's trainable parameters that are modified after each iteration. Let the 3-dimensional matrix $X^l \in \mathbb{R}^{M^l \times N^l \times C^l}$ be the input of the $l$–th convolutional layer and $G \in \mathbb{R}^{m \times n \times c^l \times S}$ be a 4-dimensional matrix, representing the $s$ kernels of $l$–th layer, of spatial span $m \times n$. The output of the $k$ – th convolutional layer will be a 3-dimensional matrix defined as $Y^l \in \mathbb{R}^{M^l - m + 1 \times N^l - n + 1 \times S}$. This matrix results from the equation

$$y_{i^l, j^l, s} = \sum_{i=0}^{m} \sum_{j=0}^{n} \sum_{k=0}^{c^l} G_{i,j,c^l,s} \times x^l_{i^l, j^l, k}. \tag{1}$$

Relation (1) is applied for all $0 \leq s \leq S$ and for any spatial span satisfying $0 \leq i^l \leq M^l - m + 1$ and $0 \leq j^l \leq N^l - n + 1$. Let $X^l \in \mathbb{R}^{M^l \times N^l \times D^l}$ be the input of the $l$–th layer that is now a pooling layer of size $n \times m$. We assume that $n$ divides $M$ and $m$ divides $N$ and the stride equals the spatial span. The output is a matrix $Y^l \in \mathbb{R}^{M^{l+1} \times N^{l+1} \times C^{l+1}}$, where

$$M^{l+1} = \frac{M^l}{n}, \quad N^{l+1} = \frac{N^l}{m}, \quad C^{l+1} = C^l, \tag{2}$$

while the polling layer operates upon $X^l$ dimension by dimension. In our network we utilize 2 max pooling layers, resulting in outputs produced based on

$$y_{i^l, j^l, d} = \max_{0 \leq i \leq m, 0 \leq j \leq n} x^l_{i^l \times n + i, j^l \times m + j, d}, \tag{3}$$

where $0 \leq i^l \leq M^l$, $0 \leq j^l \leq N^l$ and $0 \leq c \leq C^l$.

On the other hand, the fully connected layers constitute the second part of the convolutional neural network, aiming to efficiently select the most valuable features extracted from the convolutional layers. Noteworthy transition operations that connect the above layers are the batch normalization and ReLU operations. The rectifier function is described as

$$y_{i,j,d} = \max(0, x^l_{i,j,c}) \tag{4}$$

with $0 \leq i \leq M^l, 0 \leq j \leq N^l$ and $0 \leq c \leq C^l$. The input matrix $X_i$ corresponding to the $i$ −th medical image is passed through the set of successive layers and a label $\hat{y}_i$ is produced. Then an error is calculated using a defined loss function. In most cases,

Cross-Entropy loss is utilized, which is denoted as $L_{CE}$. In our occasion, where we train our network in binary classification scenarios, we utilize the Binary Cross-Entropy loss function ($L_{BCE}$), which is defined as

$$L_{BCE}(y_i, \hat{y}_i) = -(y_i * log(\hat{y}_i) + (1 - y_i) * log(1 - \hat{y}_i)) \qquad (5)$$

where $y_i = \{0, 1\}$ corresponds to the image's ground truth. The produced error is then utilized in the learning procedure that represents the modification of the *trainable* parameters of the network based on an optimization algorithm. The majority of the analyses in the literature use Adam or AdamW algorithms as optimizers.

### 3.2 Datasets

The brain cancer dataset contains 3000 images that can be utilized for the training/test of the proposed CNN architecture. The dataset is highly balanced, where 1500 images correspond to normal and 1500 images to tumorous cases. It is an open access dataset and has been uploaded on Kaggle[1].

The dataset containing lung tumors was organized by specialists in the Iraq-Oncology Teaching Hospital/ National Center for Cancer Diseases (IQ-OTH/NCCD)[2]. The dataset contains 1097 labeled images, consisting of 416 normal, 120 benign and 561 malignant cases. Since our problem is a binary-classification problem of detecting tumorous regions, we labeled as abnormal (or tumorous) all instances where benign or malignant cancer was present.

The kidney cancer dataset[3] was retrieved from a collection of cancer related datasets, which contains cases of brain, breast and other types of cancer. It contains 10000 images corresponding to images of 5000 normal and 5000 of tumorous cases, representing a completely balanced set.

### 3.3 Augmentation Pipeline

Different image-based operations are used as part of the proposed augmentation pipeline. These are *gaussian blur*, small modifications of the contrast, hue, brightness and zoom of each image (*color jittering*) as well as *rotation* and *translation*. *Gaussian Blur* is used as a cleaning mechanism of the image since it removes high frequencies in regions of the image and is commonly used as a denoising tool. Random *resize* ensures that the model will focus on infectious regions of the image independently of its height or width. Simultaneously, *rotation* and *translation* push the model to search for the tumorous region in different areas of the image and finally *color jittering* is used in order to help the model learn features that are not dependent on the color of each image, since there is also a variation in the pixel values among the grayscale images, but mainly on the shape of infectious and healthy regions. After the augmentation procedure, the MRI and CT images displaying brain, kidney and lung tumors are cropped automatically, aiming to remove their outer black parts that do not represent valuable information for the examined phenomenon.

---
[1] https://www.kaggle.com/datasets/ahmedhamada0/brain-tumor-detection
[2] https://www.kaggle.com/datasets/adityamahimkar/iqothnccd-lung-cancer-dataset
[3] https://www.kaggle.com/datasets/obulisainaren/multi-cancer

### 3.4 CNN Architecture

In this paragraph, we present a low-complexity CNN scheme including 7 main layers. The first four (2 convolutional and 2 max pooling) contribute to the feature extraction process. In addition, the 3 remaining fully connected layers take advantage of the extracted features to achieve noteworthy classification performance (Fig. 1). Two-dimensional gray-scale images of size 100×100 are placed as inputs to the proposed CNN architecture. We arrived at this choice after extensive experimentation, since the selected small input size, is accompanied by low computational burden without deteriorating the model's efficacy.

Firstly, a convolutional layer consisting of 32 kernels of 9×9 spatial span is encountered, while the extracted feature maps go through a max pooling layer of size 4×4. This pattern is replicated and contains a 5×5 convolutional and a 4×4 max pooling layer. In both cases, *same padding* is utilized before the implementation of the convolution, while the above layers are accompanied by ReLU and batch normalization operations. The second part consists of 3 fully connected layers including 4096, 1024 and 1 node, correspondingly. Between the first, exist a dropout operation aiming to eliminate the likelihood of overfitting.

## 4 Results

The proposed low-complexity scheme is employed to investigate the overall classification efficiency on all 3 datasets containing lung, kidney and brain tumors. For each dataset we implemented a stratified splitting strategy of ratio 70:30, before and after the implementation of the augmentation methodology to maintain balanced classes during the training and test phases. Thus, we result into test sets containing 900, 1500 and 329 images for the brain, kidney and lung datasets correspondingly.

Several learning rates were utilized during the training phase, namely $\eta = \{0.0001, 0.0005, 0.001, 0.005, 0.01, 0.05, 0.1\}$, while the best classification results are generated for $\eta = 0.005$. The training process has been implemented for 50 epochs using the Adam, providing a less smooth but more efficient training procedure compared to other optimizers, like the stochastic gradient descent (SGD).

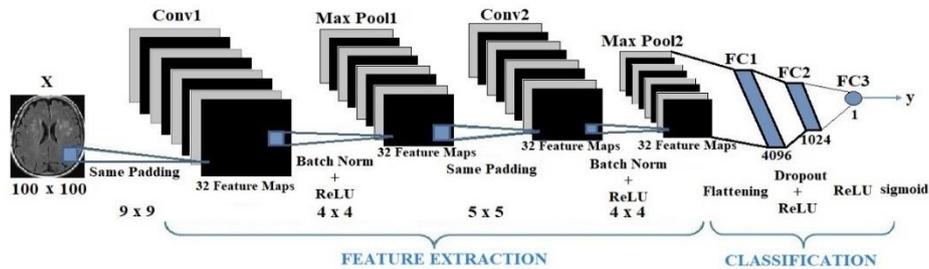

**Fig. 1.** Diagrammatic representation of the low-complexity CNN scheme

The augmentation methodology has been applied to the training set only. Since the lung cancer dataset includes 416 normal and 681 abnormal images, the application of the augmentation step helps to balance the 2 classes studied. For kidney and brain tumors this is not a concern, as these sets already have a 50:50 ratio between the 2 examined categories.

In Table 1 we observe the classification efficacy of the presented CNN on the previously mentioned test sets. Also, Table 2 displays our model's capacity after the employment of the augmentation pipeline. In both cases, our model presents significant tumor detection efficacy, especially for the augmented datasets, where we encounter accuracies of 99.33%, 100% and 99.7% for the brain, kidney and lung tumors respectively. Moreover, according to tables 1 and 2, kidney tumors represent the best classified type of tumors, before and after the employment of augmentation. Another important observation is the minor differentiation between the evaluation measures of specificity and recall, leading us to the conclusion that our model classifies equivalently effectively both tumorous and non-tumorous instances, regardless the cancer type.

**Table 1.** Classification performance of the low-complexity CNN based on the 3 examined datasets without augmentation

|  | Accuracy | Specificity | Recall | F1 score |
|---|---|---|---|---|
| **Brain Tumors** | 98.44% | 99.09% | 97.88% | 99.89% |
| **Kidney Tumors** | 99.78% | 96.55% | 99.61% | 99.78% |
| **Lung Tumors** | 97.27% | 99.11% | 96.76% | 98.05% |

**Table 2.** Classification performance of the low-complexity CNN based on the 3 examined datasets with augmentation

|  | Accuracy | Specificity | Recall | F1 score |
|---|---|---|---|---|
| **Brain Tumors** | 99.33% | 98.88% | 99.78% | 99.34% |
| **Kidney Tumors** | 100% | 100% | 100% | 100% |
| **Lung Tumors** | 99.70% | 100% | 99.49% | 99.74% |

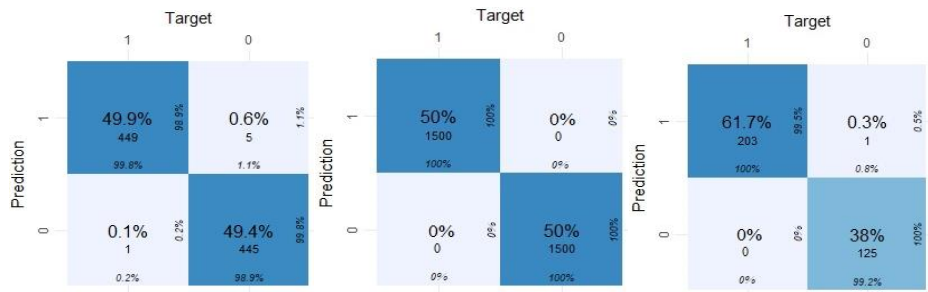

**Fig. 2.** Confusion matrices of the CNN accuracy for brain, kidney and lung images

The augmentation procedure has improved the model's overall classification efficacy in all 3 investigated scenarios, a fact that is supported by all 4 evaluation measures. More specifically, the testing accuracy is increased by 0.89% for brain tumors, 0.24% for kidney tumors, while the most prevalent increase is evident for lung cancer with an increase of 2.43%. In addition, we observe notable increases in the recall and the specificity regarding kidney and lung tumors of 2.73% and 3.44%, respectively.

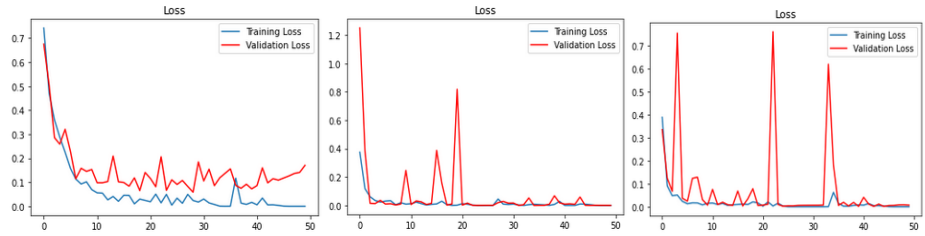

**Fig. 3.** Training/test losses for the brain, kidney and lung cancers during 50 epochs.

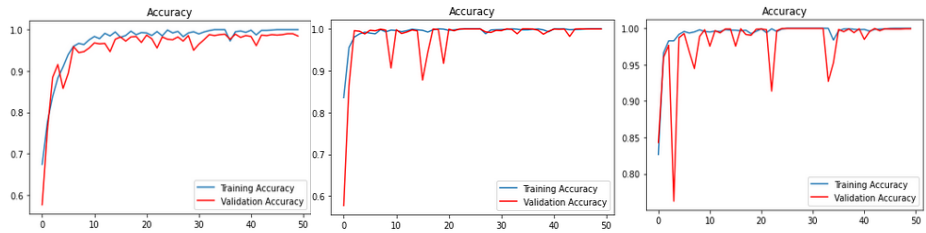

**Fig. 4.** Training/test accuracies for the brain, kidney and lung cancers during 50 epochs.

More information about the amount of correctly classified instances of each class is provided by the above confusion matrices (Fig. 2). These 3 matrices correspond to the presented results of table 2, concerning the case of the augmented training set. Finally, the diagrams of Fig. 3 and 4 display the evolution of the training/test losses and accuracies during the 50 epochs, showing a quite smooth training process, validating the robustness of the proposed AI model.

## 5    Discussion

In this article, we present a convolutional neural network scheme of low complexity for tumor anomaly detection. The limited complexity of our network renders it ideal for studying datasets including limited number of observations, that usually accompany medical studies, compared to many papers in the literature that use more complex and computationally expensive schemes. CNNs with more convolutional and more complex fully connected layers were tested, although without improving detection accuracy. Several augmentation techniques have been applied, including Gaussian blur, color jittering, rotations, resizing and translations, while we examine their influence on the detection capability of our deep learning model. The variety of datasets and the diversity of classification challenges, enhance the robustness of the presented detection efficacy, rendering our approach a reliable medical tool.

We showed that the proposed low-complexity convolutional neural network is able to produce notably accurate results based on the 4 examined classification metrics used throughout the analysis. The selected augmentation pipeline forces the detector to learn features that focus on the tumorous regions which are independent from pixel-related features, such as the brightness or hue. The increase in the evaluation metrics is obvious, resulting in accuracy scores of 99.33%, 100.00% and 99.70% for the brain, kidney and lung cancer, respectively. On top of that, since the cardinality of the medical-related datasets is mostly limited, deep architectures may fall under the threat of overfitting [30-31]. Other interesting approaches that used machine learning algorithms like random forests [22] or more complex AI schemes [12-17, 23-24, 28] generated robust results with the highest testing accuracy being 98%. Our low-complexity network coupled with the data augmentation overcomes this challenge and can be easily implemented under real-world circumstances, while its simplistic architecture encourages its retraining when new data is presented.

Since new types of cancer may occur, scientists should focus more on the generalization of their methods or the construction of architectures that will be able to provide robust results with limited amounts of data. Regarding future work, our objective is to take advantage of the knowledge of deep pre-trained networks, such as ResNet50 on ImageNet, in order to create pipelines based on transfer learning and related vision techniques. It would be interesting to apply a teacher–student method to approach the problem under the few-shot or no-shot learning scenario. Finally, it would be interesting to examine cancer occurrence simultaneously with the emergence of other chronic diseases, through causality or correlation measures [32, 33], which may lead to more precise treatment approaches.